\documentclass[aps,prl,reprint,superscriptaddress,showpacs,floatfix]{revtex4-1}
\usepackage{amsmath,amssymb}
\usepackage{graphicx}
\usepackage{dcolumn}
\usepackage{bm}
\usepackage{multirow}

\begin{document}
\title{Quantum Diffusion on Molecular Tubes: Universal Scaling of the 1D to 2D Transition}
\author{Chern Chuang}
\affiliation{Department of Chemistry, Massachusetts Institute of Technology, MA 02139, USA}
\author{Chee Kong Lee}
\affiliation{Department of Chemistry, Massachusetts Institute of Technology, MA 02139, USA}
\author{Jeremy M. Moix}
\affiliation{Department of Chemistry, Massachusetts Institute of Technology, MA 02139, USA}
\author{Jasper Knoester}
\affiliation{Zernike Institute for Advanced Materials, University of Groningen, Nijenborgh 4, AG Groningen 9747,
The Netherlands}
\author{Jianshu Cao}
\email{jianshu@mit.edu}
\affiliation{Department of Chemistry, Massachusetts Institute of Technology, MA 02139, USA}

\begin{abstract}
The transport properties of disordered systems are known to depend critically on dimensionality. We study the diffusion coefficient of a quantum particle confined to a lattice on the surface of a tube, where it scales between the 1D and 2D limits. It is found that the scaling relation is universal and independent of the disorder and noise parameters, and the essential order parameter is the ratio between the localization length in 2D and the circumference of the tube. Phenomenological and quantitative expressions for transport properties as functions of disorder and noise are obtained and applied to real systems: In the natural chlorosomes found in light-harvesting bacteria the exciton transfer dynamics is predicted to be in the 2D limit, whereas a family of synthetic molecular aggregates is found to be in the homogeneous limit and is independent of dimensionality.
\end{abstract}

\maketitle

\textit{Introduction.}---Transport of energy or charge carriers is of fundamental importance in terms of both scientific interest and its technological relevance. The seminal work of Anderson states that the presence of static disorder leads to a metal-to-insulator transition or even totally prevents transport in lower dimensions\cite{Anderson1958,Anderson1979}. Upon coupling to fluctuating environment, localized quasi-particles can overcome energetic barriers, and the system becomes conductive again\cite{Thouless1981}. While transport ceases to exist in both the zero coupling limit (Anderson localization) and the strong coupling limit (dynamical localization), the intervention of environmental noise with intermediate strength can maximize the conductivity\cite{Wolynes1990,Leitner1996,JeremyNJP}.

Compared to classical hopping kinetics, where the governing rate equations are given in the coordinate basis, the motion of quantum particles on a disordered and noisy lattice is more involved. In fact, in the weak system-environment coupling limit, the dynamics of the particle wavefunction can be cast into rate equations in the eigenbasis. This implies that \textit{quantum enhancement} of the conductivity can be characterized by the average size of the wavefunctions, the localization length, since this corresponds to the step size of each hopping event\cite{JeremyNJP,CheeKong}. An immediate consequence arises if one considers the different scaling behaviors of the localization length in different dimensions. It is expected that, for example, the quantum enhancement is much stronger in 2D with respect to that in 1D, given the same disorder and noise strength. 

In this Letter we investigate the diffusive dynamics of a quantum particle on a tubular lattice in the axial direction, in which the transport properties scale between the 1D and the 2D limits. Recently, the optical and dynamic properties of excitons in natural\cite{deGroot2009,JPCL.3.3745,JACSChlorosome2011} and synthetic\cite{ChemRev2014,JPCB.113.2273,JPCB.101.5646,Dorthe2012,JPCC.117.26473,PRL.108.067401} self-assembled tubular molecular aggregates have drawn much attention. The combination of their quasi-one-dimensional (wire-like) structure and the attenuation of exciton localization due to their inherent (locally) 2D nature, makes such tubular aggregates potentially ideal for exciton transport in, for instance, photovoltaic devices\cite{JasperPRL2015}. A natural order parameter in this regard is the radius of the tube, where the axial conductivity is found to be an increasing function of the radius until a critical radius is reached and levels off as it approaches the 2D limit. We found that the scaling relation is universal, independent of the parameters chosen. Moreover, the critical radius is shown to be directly proportional to the localization length in the corresponding 2D system. A phenomenological expression is proposed and shown to reproduce the radius dependence quantitatively, which is applied to several real systems in different limiting parameter regimes and predict their respective radius-(in)dependent diffusion constant.

\textit{Calculation of quantum diffusion.}---The Haken-Strobl-Reineker (HSR) model is employed to characterize the system of interest coupled to a classical Markovian noisy environment\cite{KenkreReineker,Madhukar1977,Amir2009}. The dynamics of the system is described by the stochastic Schr\"{o}dinger equation
\begin{eqnarray}
i\frac{d}{dt}|\psi\rangle = \hat{H}_\mathrm{s}|\psi\rangle + \sum_\mathbf{n} F_\mathbf{n}(t)\hat{V}_\mathbf{n}|\psi\rangle,\label{eqn:StoSchro}
\end{eqnarray}
where $\hat{V}_\mathbf{n}=|\mathbf{n}\rangle\langle\mathbf{n}|$, $F_\mathbf{n}(t)$ are Gaussian stochastic processes with zero mean ($\langle F_\mathbf{n}(t)\rangle=0$) and finite second order autocorrelation $\langle F_\mathbf{n}(t)F_\mathbf{m}(s)\rangle=\Gamma\delta_\mathbf{nm}\delta(t-s)$, with $\Gamma$ the dephasing rate. The system Hamiltonian $\hat{H}_\mathrm{s}$ is characterized by a nearest-neighbor coupled square-lattice with periodic boundary condition in one direction (circumference) and isotropic coupling constant $J$. The number of sites ($R$) along the tube's circumference is referred to as the radius of the tube. The energy of site $\mathbf{n}$, $\epsilon_\mathbf{n}$, is taken to be an independent Gaussian random variable with standard deviation $\sigma$.

The central physical observable in this Letter, the diffusion coefficient $D$ in the direction given by unit vector $\vec{u}$, is given by the Green-Kubo expression,
\begin{eqnarray}
D(\mathbf{u})=\frac{1}{Z_\mathrm{s}}\int_0^\infty dt\mathrm{Tr}\left[e^{-\beta \hat{H}_\mathrm{s}}\hat{j}(\mathbf{u},t)\hat{j}(\mathbf{u})\right],\label{eqn:GreenKubo}
\end{eqnarray}
where $Z_\mathrm{s}$ is the system partition function. In the context of the HSR model we will take $\beta=0$ (infinite temperature), so $Z_\mathrm{s}=N$, where $N$ is the size of the system. Moreover, we show that the time integration can be carried out analytically.

\begin{eqnarray}
D(\mathbf{u})=\frac{1}{N}\sum_{{\mu},{\nu}=1}^N\frac{\Gamma}{\Gamma^2+\omega_{{\mu}{\nu}}^2}|\hat{j}_{\mu\nu}(\mathbf{u})|^2,
\label{eqn:GreenKuboD}
\end{eqnarray}
where $\hat{j}_{\mu\nu}(\mathbf{u})$ is the flux operator in the eigenbasis and $\omega_{\mu\nu}=\omega_\mu-\omega_\nu$ is the difference between the energies of states $\mu$ and $\nu$, with details recorded in the Supplemental Material. It typically takes up to 100 sites in the axial direction to converge the results for the range of disorder strength covered in this Letter. The diffusion coefficient obtained through Eq.~(\ref{eqn:GreenKuboD}) is quantitatively agreeing with that from propagating Eq.~(\ref{eqn:StoSchro}) as was done in Ref.~\onlinecite{JeremyNJP}. For consistency we present exclusively the data obtained with Eq.~(\ref{eqn:GreenKuboD}) in this Letter. An efficient method of propagating Eq.~(\ref{eqn:StoSchro}) in the weak coupling regime ($\Gamma/J\ll1$) is also presented in the Supplemental Material. The same methodology is applicable to the case where the system is weakly coupled to a real quantum bath in the low temperature regime, as elaborated in later sections.

The present model is exactly solvable in some limiting cases. Firstly, when the system is homogeneous, \textit{i.e.}~$\sigma=0$, the full dynamics of the quantum particle can be resolved and shows transient ballistic behavior before transitions to long-time diffusive motion\cite{Madhukar1977,JeremyNJP}. In particular, there is no dependence on the dimensionality, and the diffusion coefficient is given by\cite{CheeKong}
\begin{eqnarray}
D_\mathrm{hom}=2J^2/\Gamma,
\label{eqn:DHomogeneous}
\end{eqnarray}
which can be obtained by assuming Bloch wavefunctions $\phi_m^\mu=\exp(i\mu m)/\sqrt{N}$ in Eq.~(\ref{eqn:GreenKuboD}). In fact, decoupling of directions is valid as long as the wavefunctions of the system can be factorized into contributions of different dimensions: $\Psi(\vec{n})=\psi(n_1)\psi(n_2)\cdots\psi(n_M)$. One such example is given by considering stacks of homogeneous rings with energy bias among different rings\cite{SpanoJCP1997,Chuang2014}. In the opposite extreme where either disorder ($\sigma/J\gg1$) or system-environment coupling ($\Gamma/J\gg1$) is large, all quantum coherence is destroyed. The particle behaves classically and can be described by a hopping rate between connected sites\cite{CaoSilbey2009,NJP.12.105012,PRL.110.200402}
\begin{eqnarray}
D_\mathrm{hop}=\frac{2J^2\Gamma}{\Gamma^2+\sigma^2}.
\label{eqn:DHopping}
\end{eqnarray}
Since the hopping events are independent along different directions, independence on dimensionality is also expected in this case. Detailed derivation of Eqs.~(\ref{eqn:GreenKuboD}) and (\ref{eqn:DHomogeneous}) is given in the Supplemental Material. Hence, we conclude that prominent radius dependence of the diffusive behavior is expected only if the wavefunctions are non-separable and with finite noise strength.

In the weak damping regime with finite disorder, through a scaling argument, one can show that the diffusion coefficient can be estimated by 
\begin{eqnarray}
D_\mathrm{coh}=\Gamma\xi^2,
\label{eqn:Dcoherent}
\end{eqnarray}
where $\xi$ is the localization length of the system. This relation is very useful since it connects the dynamical observable (steady-state diffusion coefficient) with a static property of the system and a single parameter characterizing the system-environment coupling\cite{Wolynes1990,JeremyNJP}. We provide a heuristic derivation of this relation for 1D systems in the Supplemental Material while its applicability will be demonstrated in the following section. 

\begin{figure}[t]
	\centering
  \includegraphics[width=8cm]{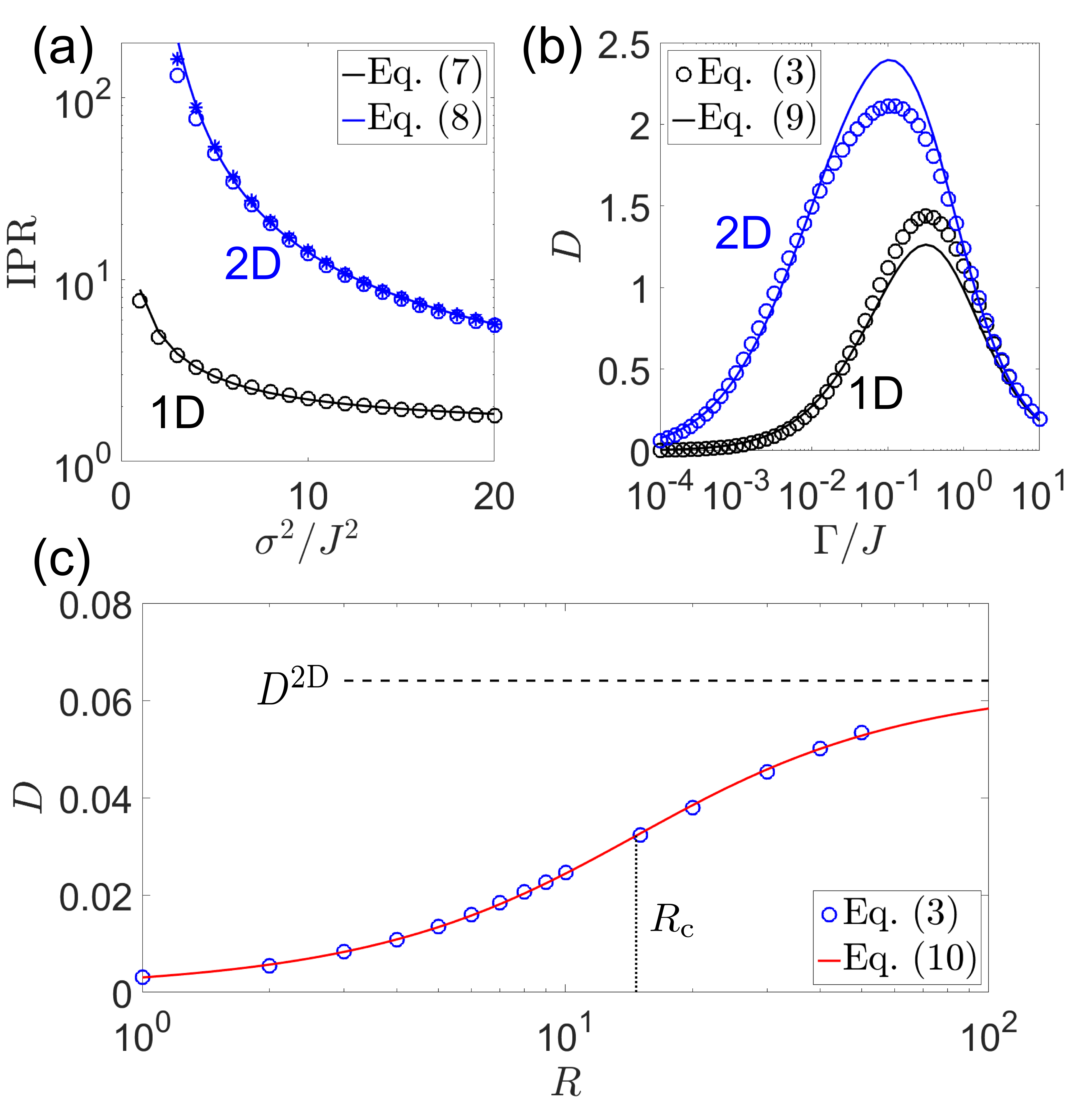}
  \caption{(Color online) (a) IPR dependence on the disorder strength in 1D and 2D. The parameters fitted in Eqs.~(\ref{eqn:IPR1D}) and (\ref{eqn:IPR2D}) are $a_1=6.2$ and $(a_2,b_2)=(67,6.7)$. The numerical data are shown in symbols. We use 4900 sites for 1D system (black circles), and 2D systems with 70$\times$70 (blue circles) and 90$\times$90 (blue asterisks) square lattices (b) Comparison between the numerically exact results of Eq.~(\ref{eqn:GreenKuboD}) and the diffusion coefficients obtained from the interpolation formula~(\ref{eqn:Thouless}). The lower (black) circles and solid line refer to 1D systems and the upper (blue) circles and line represent 2D systems. In both cases we set $\sigma/J=1$. (c) Radius dependence of $D$ in the Redfield regime ($\Gamma/J=10^{-4}$) with $\sigma/J=1$. The solid line is the fitting according to Eq.~(\ref{eqn:Rc_fit}), with the corresponding fitted parameters $R_\mathrm{c}$ and $D^\mathrm{2D}$ indicated.}
  \label{fig:D1D2D}
\end{figure}

\textit{Numerical results.}---We start by discussing the diffusion constants in 1D and 2D. It is well known that the scaling of the localization length depends critically on dimensionality; specifically it scales linearly with the mean free path in 1D and exponentially in 2D\cite{LeeRMP1985}. A common and useful measure of the localization length is given by the inverse participation ratio (IPR), defined for each of the eigenstates as $\mathrm{IPR}_\mu=1/\sum_\mathbf{m}|\phi_\mathbf{m}^\mu|^4$. Due to the high temperature characteristic of the HSR model we average over all eigenstates and fit the IPR of disordered 1D and 2D square lattices according to
\begin{eqnarray}
\xi^\mathrm{1D}&=&\mathrm{IPR}^\mathrm{1D}\sim a_1l,\label{eqn:IPR1D}\\
(\xi^\mathrm{2D})^2&=&\mathrm{IPR}^\mathrm{2D}\sim a_2l\exp\left(b_2l\right)\label{eqn:IPR2D},
\end{eqnarray}
where $l=\frac{J^2}{\sigma^2}$ is the mean free path, and length scale is measured in units of the lattice constant. The results are shown in Fig.~\ref{fig:D1D2D}(a), with fitted parameters indicated in the caption. These expressions provide a simple way of estimating the diffusion coefficient in the weak damping regime given the disorder strength $\sigma$, where Eq.~(\ref{eqn:Dcoherent}) applies. Note that in 1D the IPR is directly interpreted as the localization length, while in 2D its square root is. This is because it is the diffusion along one particular direction that concerns us. 

Thouless and Kirkpatrick proposed an interpolating formula for the general case which was proven to be valid for most of the parameter ranges of interest\cite{Thouless1981,JeremyNJP}:
\begin{eqnarray}
D_\mathrm{interp}=\left[\left(\frac{2J^2}{\Gamma+\sigma/2}\right)^{-1/2}+(\Gamma\xi^2)^{-1/2}\right]^{-2}.
\label{eqn:Thouless}
\end{eqnarray}
In Fig.~\ref{fig:D1D2D}(b) this interpolation result is shown as a function of $\Gamma$ and compared to the numerically exact results obtained from Eq.~(\ref{eqn:GreenKuboD}), averaging over 100 realizations of disorder. At a given disorder strength $\sigma$ one expects an optimal dephasing rate promotes maximal diffusive transport\cite{JeremyNJP,LiamNJP}. The interpolation formula not only describes the two limits correctly, but also captures the maxima almost quantitatively, showing the transition between the two transport mechanisms. Note that this expression also reproduces the convergence of diffusion constants in different dimensions in the homogeneous limit, \textit{i.e.}, Eq.~(\ref{eqn:DHomogeneous}).

\begin{figure}[t]
	\centering
  \includegraphics[width=8cm]{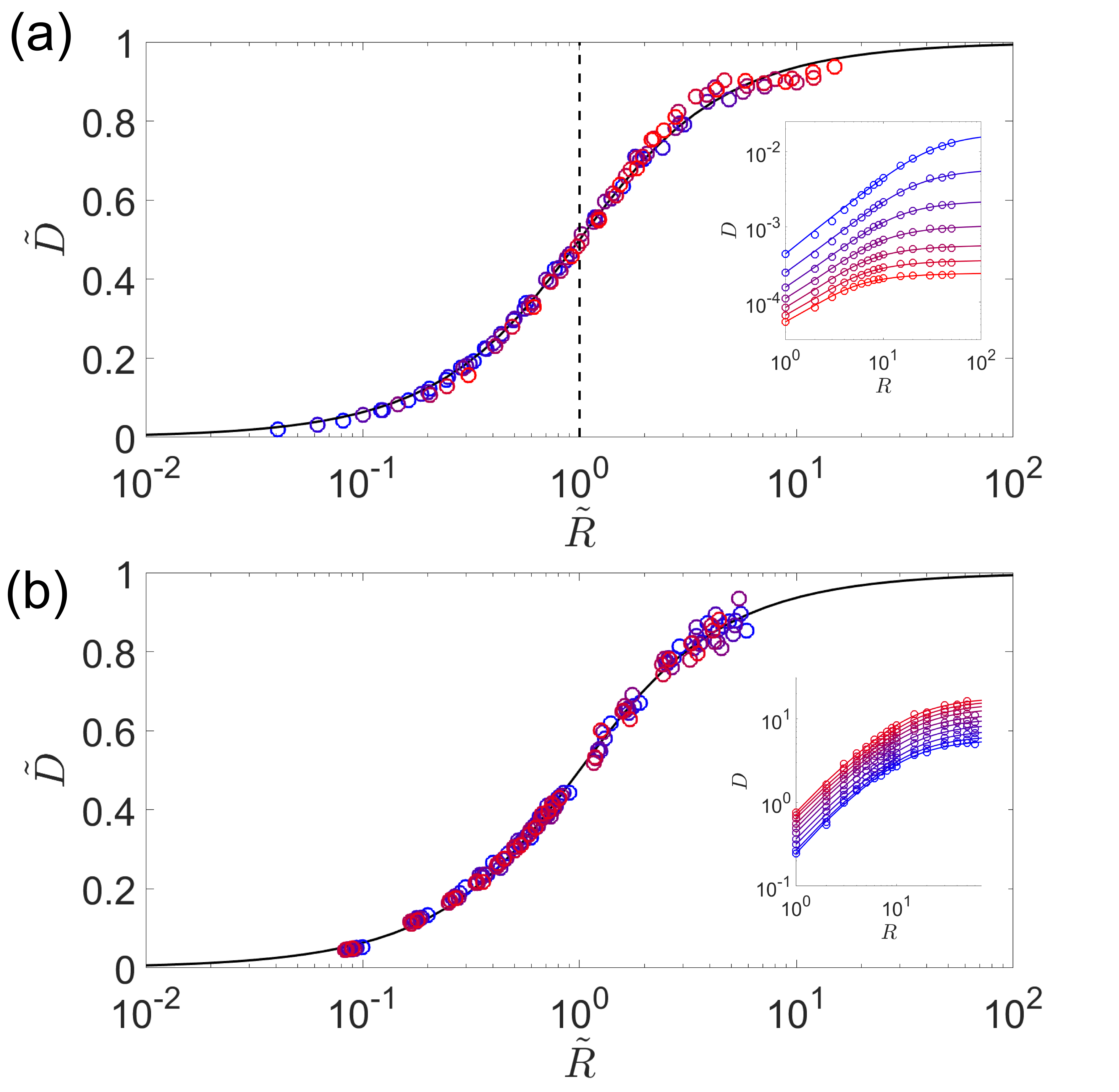}
  \caption{(Color online)  (a) Relative diffusion coefficient $\tilde{D}=(D(R)-D^\mathrm{1D})/(D^\mathrm{2D}-D^\mathrm{1D})$ as a function of rescaled radius $\tilde{R}=(R-1)/R_\mathrm{c}$. The solid line is the fitting function $S(x)$ and the dashed line indicates $\tilde{R}=1$. The inset shows the original data before rescaling: From top ($\sigma/J=2$, blue) to bottom ($\sigma/J=5$, red) with $0.5$ increment and interpolating color gradient. (b) Results of the secular Redfield calculations with varying temperatures, with the same convention used in (a). Inset: From top ($T/J=7$, red) to bottom ($T/J=0.7$, blue) with $0.7$ decrement.}
  \label{fig:DofR}
\end{figure}

We next look at the radius dependence of the diffusion constant in a tube. Since the diffusive motion in the large dephasing limit does not depend on system dimension and the $\Gamma$ dependence is well described by Thouless's interpolation formula, it is instructive to focus on the Redfield regime ($\Gamma/J\ll1$) when investigating the radius dependence. (Further analysis on the effect of finite $\Gamma$ is provided in the Supplemental Material.) This dependence should be bounded from below by the results of 1D diffusion and from above by 2D diffusion, as seen in Fig.~\ref{fig:D1D2D}(c). Here the diffusion constant increases as the tube radius $R$ increases until the trend is attenuated at the inflection point $R=R_\mathrm{c}$, denoted as the critical radius. Moreover, we found this radius dependence to be universal across the entirety of the parameter space we scanned, as shown in Fig.~\ref{fig:DofR}(a), where we have rescaled the data according to the following phenomenological expression
\begin{eqnarray}
D(R)&=&D^\mathrm{1D}+(D^\mathrm{2D}-D^\mathrm{1D})S\left(\frac{R-1}{R_\mathrm{c}}\right),
\label{eqn:Rc_fit}
\end{eqnarray}
where $S(0)=0$, $S(\infty)=1$, and $dS/dx$ is everywhere positive for $x>0$. Here we chose $S(x)=2\arctan(x)/\pi$.\footnote{The numerical evaluation of Eq.~(\ref{eqn:GreenKuboD}) becomes demanding for 2D lattices with weak disorder. In the $R$-dependence data presented the number of sites in the axial direction is fixed to 100, with sites on the circumference up to 50. We then fit both $D^\mathrm{2D}$ and $R_\mathrm{c}$ according to Eq.~(\ref{eqn:Rc_fit}).} To demonstrate the generality of this observation, we also present the universality found for the quantum diffusion with realistic quantum bath treated with secular Redfield approximations in Fig.~\ref{fig:DofR}(b). With the details described in the Supplemental Material, this method models accurately the low temperature thermal activated transport regime that complements the HSR model\cite{BasslerPRB,BasslerReview,CheeKong}, and has been shown to explain the temperature dependent exciton properties of molecular aggregates relevant to our discussion in the next section\cite{Jasper2003PRL,Jasper2005PRL}. 

\begin{figure}[t]
	\centering
  \includegraphics[width=7cm]{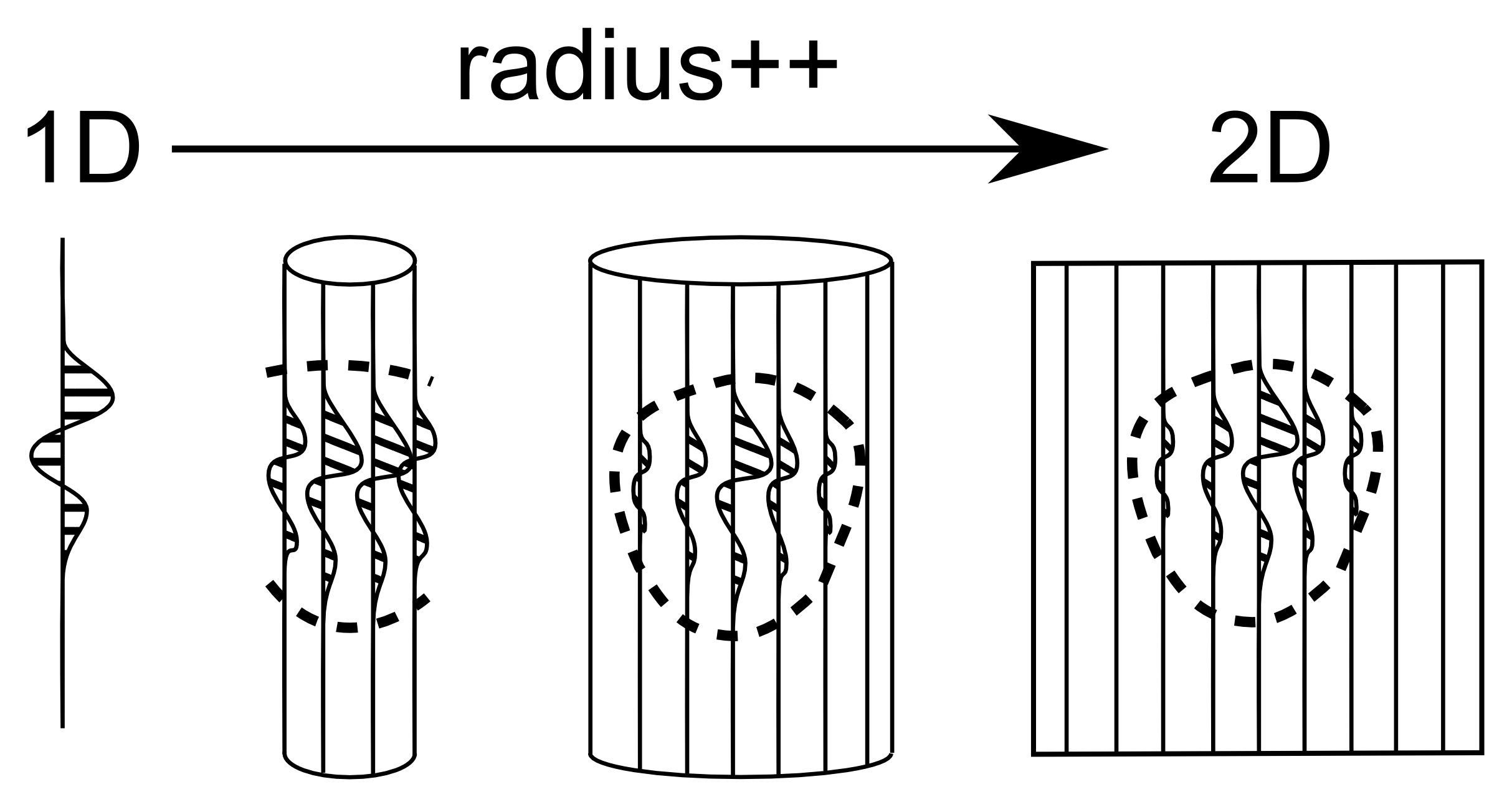}
  \caption{Schematic illustration of the origin of universal radius scaling of transport rate in tubes. Interpolating between the 1D and the 2D limits, the localized wavefunctions on a tubular lattice no longer encompass the tube as the radius exceeds the critical value and vice versa.}
  \label{fig:Schematics}
\end{figure}

The universality can be explained by the following interpretation. One expects a strong radius dependence of the diffusion coefficient only if the particle wavefunction fully delocalizes around the tube. This is no longer valid as the radius becomes larger than its critical value, where the wavefunction occupies only partially the space in the circumferential direction. Essentially, this picture identifies the critical radius with the localization length in the corresponding 2D system, as is illustrated in Fig.~\ref{fig:Schematics}. In determining the radius dependence, one compares two length scales of the system: the circumference of the tube and the inherent localization length along the circumference. This makes our theory predictive on the axial diffusion coefficients of general tubular systems, given the knowledge of the localization length obtained from experiments or \textit{ab initio} calculations, as demonstrated in the next section. We note that this picture can also be applied to understanding the optical properties of tubular systems. Specifically, it has been shown that the existence of linear dichroism signals depends critically on the ratio of radius and localization length\cite{JCP.134.114507}. 

\textit{Estimates for real systems.}---Adding to the previous generic analysis, it is informative to analyze real systems that have been characterized experimentally. There are numerous examples of tubular aggregates consisting of organic chromophores. Amongst the best known examples found in nature are the chlorosomes in green sulfur bacteria\cite{deGroot2009}, which serve as the antenna of the light-harvesting apparatus. Diffusive transport of excitons in chlorosomes has been identified\cite{JACSChlorosome2012,Aspuru2012}. The above scaling argument predicts the exciton diffusion on chlorosome tubes to be in the 2D limit, because the critical radius $R_\mathrm{c}$ is much smaller than the typical radius found in the organism, see Table~\ref{tab:RealSys}\footnote{In real systems, the interactions are of long-range and anisotropic dipolar character. Here, we use an effective nearest-neighbor coupling by demanding that the total bandwidth equals that of a 2D dipole lattice. The dephasing rate $\Gamma$ is estimated by taking the fast bath limit of a real quantum bath such that the bath correlation function becomes a $\delta$-function, \textit{e.g.} we use $\Gamma\approx2\lambda/\beta\omega_\mathrm{c}$ for a Drude-Lorentz bath $J(\omega)=2\lambda\omega_\mathrm{c}\omega/(\omega_\mathrm{c}^2+\omega^2)$}. This implies the chlorosomes fully exploit the enhancement and robustness of quantum transport in 2D compared to 1D, while taking the advantage of broad absorption spectrum induced by strong homogeneous ($\Gamma$) and inhomogeneous ($\sigma$) broadening mechanisms\footnote{We note that the chlorosome tubes lie \textit{horizontally} on the baseplate, below which the FMO complexes and the reaction centers are located. Our argument of the 2D enhancement of quantum diffusion applies to the transport along the circumferential direction as well. In addition, when photon influx is large the excitons can migrate in the axial direction and eventually reach unoccupied reaction centers.}.

Families of synthetic tubular molecular aggregates exist as well\cite{Didraga2004,JPCC.118.24854,ChemRev2014}. Similar to chlorosomes found in nature, most of them comprise amphiphilic molecules self-assembled into (multi-layer) nanotubes with length up to micron scale. One such molecular aggregate, composed of the dye molecule C8S3, has been recently characterized\cite{Dorthe2012,JPCL.5.2274} and comprises two concentric nanotubes. In this case, due to the reduced static disorder and strong exciton coupling strength ($J\approx8\sigma$, see Table~\ref{tab:RealSys}), the system is in the homogeneous limit and the diffusion coefficient becomes independent of dimensionality and thus independent of radius. This conclusion is supported by the existence of well-defined absorption selection rules arising from the wavefunctions fully delocalized around the circumferences\cite{Didraga2004, Dorthe2012}. The large localization length in such systems can be utilized in transporting the excitons efficiently along the tubes\cite{JPCL.5.2274,JasperPRL2015}.

There are other instances where radius (in)dependence of transport in tubular systems exists. It has been shown that the increment of semiconducting single-walled carbon nanotube radius is accompanied by a mild linear increase of exciton mobility\cite{RiceCNT}. This implies that the reduced disorder in clean carbon nanotube samples gives rise to large localization length of the exciton wavefunction\cite{CNTbook}, so the system is in the $R\ll R_\mathrm{c}$ limit and a linear radius dependence is predicted. In addition, molecular tubes based on self-assembled tobacco mosaic virus protein monomers designed to mimic natural light-harvesting arrays were synthesized\cite{TMVJACS2007}. It is found that the exciton dynamics can be described appropriately by classical hopping kinetics\cite{KimCao2010,Whaley2013}, thus the independence of dimensionality is predicted (see Table~\ref{tab:RealSys}). Lastly, quantum diffusion of excitons in aggregated phycocyanin thin films has been experimentally characterized recently\cite{PaltielPCCP}, where the delocalization of excitons explains the enhancement of the diffusion length compared to the estimate of classical hopping theory. While this artificial system serves as an example of quantum diffusion in 2D under the current scheme, the naturally occurring form of phycocyanin in most cyanobacteria self-assembles into a finite 1D wire\cite{JPlantPhysiol.168.1473}. It is our ongoing effort to analyze this interesting system in detail in this regard. 

Both the HSR model and the weak-coupling secular Redfield method applied to isotropic nearest-neighbor coupled square lattices are an over-simplification of the real systems\cite{Note2}. Even richer physical content can be expected when considering more realistic aspects. For examples, it has been shown that environmental memory effects can enhance diffusive transport\cite{NJP.12.105012,Aspuru2012}. The anisotropy from nontrivial molecular arrangement has also been shown, for example, to render a helical character to the exciton wavefunction\cite{JCP.112.3023,JCP.134.114507,LiamPNAS,Arend2D}. In addition, the statistical nature of disorder\cite{BasLocalization} and long-range interactions\cite{MalyshevLocalization} are both critical in determining the localization length. We expect the $R$-dependence to be more involved in these and other possible generalizations, since the functional dependence of the localization length on the additional model parameters varies. However, once given these parameters and consequently the localization length, the transition from 1D to 2D can be compactly characterized by the ratio between radius and the localization length, in accordance with the physical picture discussed in the previous section. As a result, we believe the universal scaling relation investigated in this Letter can serve as a generic guidance. Finally the theoretical framework developed here also applies to, for example, the in-plane exciton mobility as a function of the thickness of thin films, which is predicted to scale between the 2D and the 3D limits.

\begin{table}
\begin{center}
\def\arraystretch{1.5}
\begin{tabular}{|c|c|c|c|c|c|c|c|}
\hline
			&	$J$		&	$\sigma$	&	$\Gamma$	&	$D^\mathrm{1D}$	&	$D^\mathrm{2D}$	&	$R_\mathrm{c}$	&	R\\ \hline
Chlorosome	&	400		&	1000		&	350			&	26				&	35				&	6				&	~100\\ \hline
C8S3 tube	&	2000		&	250		&	300			&	2800				&	2900				&	-			&	30/60\\ \hline
TMV	 tube 	&	50		&	3000		&	400			&	0.6				&	0.6				&	1				&	17\\ \hline
\end{tabular}
\end{center}
\caption{Parameters and axial exciton diffusion estimated\cite{Note2} for three exemplary real tubular systems at room temperature. $J$, $\sigma$, $\Gamma$ are given in cm$^{-1}$, diffusion coefficients are in nm$^2$/ps, and $R$ ($R_c$) is unitless representing the (critical) number of molecules in the circumference. The parameters for $J$, $\sigma$, $\Gamma$, and $R$ from top to bottom are deduced from Ref.~\onlinecite{Aspuru2012},~\onlinecite{JPCA.114.8179,Dorthe2012}, and~\onlinecite{Whaley2013}, respectively. The other quantities are calculated using Eqs.~(\ref{eqn:IPR1D}),~(\ref{eqn:IPR2D}), and~(\ref{eqn:Thouless}). We take $R_\mathrm{c}=\xi^\mathrm{2D}$ since the ratio between $\xi^\mathrm{2D}$ from Eq.~(\ref{eqn:IPR2D}) and the fitted $R_\mathrm{c}$ from Eq.~(\ref{eqn:Rc_fit}) is close to unity in our calculations. }
\label{tab:RealSys}
\end{table}%
\textit{Conclusion.}---We have developed a theoretical framework and efficient numerical procedure to model exciton dynamics in tubular molecular aggregates in the presence of environmental noise and disorder based on the HSR model and the secular Redfield model. The central observation is that the diffusion coefficient along the axial direction increases as a function of the tube radius. This dependence is found to be universal across the full parameter range of interest, and can be quantitatively defined by the ratio between the two length scales involved: the circumference of the tube and the localization length of the corresponding 2D system. For the chlorosome tubes found in green sulfur bacteria, the exciton transport is found to be in the 2D limit. On the other hand, in a synthetic system with self-assembled cyanine dye molecules mimicking chlorosomes, the excitons are in the homogeneous limit where independence of dimensionality is predicted. Our findings are useful when exploiting the structure-property relation in designing robust and efficient artificial light-harvesting devices.


\bibliography{HSR_Tube}


\end{document}


\title{Supplemental on-line material for\\
Quantum Diffusion on Molecular Tubes: Universal Scaling of the 1D to 2D Transition}
\author{Chern Chuang}
\affiliation{Department of Chemistry, Massachusetts Institute of Technology, MA 02139, USA}
\author{Chee Kong Lee}
\affiliation{Department of Chemistry, Massachusetts Institute of Technology, MA 02139, USA}
\author{Jeremy M. Moix}
\affiliation{Department of Chemistry, Massachusetts Institute of Technology, MA 02139, USA}
\author{Jasper Knoester}
\affiliation{Zernike Institute for Advanced Materials, University of Groningen, Nijenborgh 4, AG Groningen 9747,
The Netherlands}
\author{Jianshu Cao}
\email{jianshu@mit.edu}
\affiliation{Department of Chemistry, Massachusetts Institute of Technology, MA 02139, USA}

\maketitle

\setcounter{section}{0}
\renewcommand{\thesection}{S\arabic{section}}
\setcounter{figure}{0}
\renewcommand{\thefigure}{S\arabic{figure}}
\setcounter{equation}{0}
\renewcommand{\theequation}{S\arabic{equation}}

\subsection{Efficient disorder sampling in the Redfield regime: HSR model}
As mentioned in the main text, significant $R$-dependence exists only in the weak to moderate noise strength regime. In the case where $\Gamma/J\ll 1$, the system dynamics can be accurately described by the Redfield equations. Thus, the population dynamics is governed by master equations in the eigenbasis
\begin{eqnarray}
\dot{\rho}_{\mu\mu}=\sum_\nu W_{\mu\nu}\rho_{\nu\nu},
\label{eqn:RedfieldEqn}
\end{eqnarray}
with the rate matrix element given by
\begin{eqnarray}
W_{\mu\nu}&=&\Gamma\sum_n|\phi_n^\mu|^2|\phi_n^\nu|^2\label{eqn:RedfieldOffDiag}\\
W_{\mu\mu}&=&-\Gamma|\sum_n\phi_n^\mu|^2(1-|\phi_n^\mu|^2)=-\sum_{\nu\neq\mu}W_{\mu\nu}.
\label{eqn:RedfieldDiag}
\end{eqnarray}
We use Greek letters for the eigenstates of the system and Roman letters for the local basis throughout. Now, an expression for the time-dependent diffusion coefficient can be obtained straightforwardly by taking the time derivative of the mean squared displacement,
\begin{eqnarray}
D(t)&=&\sum_nn^2\sum_\mu|\phi^\mu_n|^2\sum_\nu W_{\mu\nu}\rho_{\nu\nu}(t)\label{eqn:RedfieldDiffusion}\\
&=&\frac{\Gamma}{2}\sum_{m,n,\mu}\left(m^2|\phi_m^\mu|^2|\phi_n^\mu|^2-n^2\right)\rho_{nn}(t).
\label{eqn:RedfieldDiffusionHSR}
\end{eqnarray}
Several remarks are due with regard to the derivation of the last expression. Firstly in deriving it we assume that all coherence in the eigenbasis vanishes, which is valid in the steady-state limit ($\Gamma t >1$) where diffusive dynamics dominates. In practice we propagate the system to, for example, $t=10/\Gamma$ from a $\delta$-localized initial state before any significant boundary effect arises. Second, by assuming periodic boundary condition, the same set of Redfield rate matrix elements can be used to propagate $N_\mathrm{t}=N\cdot R$ different initial conditions in a matrix multiplication fashion, where $N_\mathrm{t}$ is the total number of sites and $N$~($R$) is the axial (circumferential) dimension of the tube sample. Note that this is extremely economical numerically and is working best for when looking at large, for example, 2D systems. The result agrees quantitatively well with the Green-Kubo expression Eq.~(3) in the main text.

\subsection{Derivation of Eq.~(3)}
The flux operator $\hat{j}(\mathbf{u},t)$ and its time derivative is given by
\begin{eqnarray}
\hat{j}(\mathbf{u}) &=& i\sum_{\mathbf{n},\mathbf{m}}(\mathbf{u}\cdot\mathbf{r}_{\mathbf{n}\mathbf{m}})J_{\mathbf{n}\mathbf{m}}|\mathbf{m}\rangle\langle\mathbf{n}|,\\
\frac{d}{dt}\hat{j}(t)&=&i\left[\hat{j}(t),\hat{H}_\mathrm{s}\right]-\frac{\Gamma}{2}\sum_\mathbf{n}\left[\hat{V}_\mathbf{n},\left[\hat{V}_\mathbf{n},\hat{j}(t)\right]\right]\label{eqn:djdt}.
\end{eqnarray}

We then write the flux operator in the eigenbasis
\begin{eqnarray}
\hat{j}_{\mu\nu}(\mathbf{u})&=&i\sum_{\mathbf{n},\mathbf{m}=1}^N\left(\mathbf{\mathbf{u}}\cdot\mathbf{r}_{\mathbf{m}\mathbf{n}}\right)\phi^{{\mu}*}_\mathbf{n}\phi^{{\nu}}_\mathbf{m}J_{\mathbf{n}\mathbf{m}},
\end{eqnarray}
where $\omega_{\mu\nu}$ stands for the energy difference between eigenstates $\mu$ and $\nu$, $\mathbf{r}_{\mathbf{m}\mathbf{n}}=\mathbf{r}_\mathbf{m}-\mathbf{r}_\mathbf{n}$ is the vector connecting the two sites, and $\phi^\mu_\mathbf{m}$ is the wavefunction amplitude of state $\mu$ at site $\mathbf{m}$. 
Starting from Eq.~(\ref{eqn:djdt}), one can obtain the formal solution for the flux operator in the Heisenberg picture,
\begin{eqnarray}
\hat{j}(t)=e^{i\hat{H}_\mathrm{s}t}\hat{j}e^{-i\hat{H}_\mathrm{s}t}e^{-\Gamma t}.
\end{eqnarray}
Substituting the above expression into Eq.~(2), the Green-Kubo formula, carrying out the integration over time, and inserting two completeness relations in the system eigenbasis, we readily arrive at Eq.~(3).

\subsection{Derivation of Eq.~(4) from Eq.~(3)}
In the homogeneous limit the wavefunction is given by the Bloch function, $\phi_{m}^\mu=e^{2\pi i\mu m/N}/\sqrt{N}$. Moreover, the coupling is translationally invariant, so $J_{n,m}=J(|\Delta|)$, where $\Delta=m-n$. Thus,
\begin{widetext}
\begin{eqnarray}
D_\text{hom}&=&-\frac{1}{N^3}\sum_{\mu,\nu}\frac{\Gamma}{\Gamma^2+\omega_{\mu\nu}^2}\sum_{n,m}\sum_{n',m'}J_{n,m}J_{n',m'}(m-n)(m'-n')e^{\frac{2\pi i}{N}\left[\mu(m'-n)+\nu(m-n')\right]}\nonumber\\
&=&-\frac{1}{N^3}\sum_{\mu,\nu}\frac{\Gamma}{\Gamma^2+\omega_{\mu\nu}^2}\sum_{\Delta,\Delta'}e^{\frac{2\pi i}{N}(\mu\Delta'+\nu\Delta)}J(|\Delta|)J(|\Delta'|)\Delta\Delta'\sum_{n,n'}e^{\frac{2\pi i}{N}(n'-n)(\mu-\nu)}\nonumber\\
&=&-\frac{1}{N}\sum_{\Delta,\Delta'}J(|\Delta|)J(|\Delta'|)\Delta\Delta'\sum_{\mu,\nu}\frac{\Gamma\delta_{\mu,\nu}}{\Gamma^2+\omega_{\mu\nu}^2}e^{\frac{2\pi i}{N}(\mu\Delta'+\nu\Delta)}\nonumber\\
&=&-\frac{1}{\Gamma}\sum_{\Delta,\Delta'}J(|\Delta|)J(|\Delta'|)\Delta\Delta'\delta_{\Delta,-\Delta'}\nonumber\\
&=&\frac{1}{\Gamma}\sum_\Delta J(|\Delta|)^2\Delta^2.
\end{eqnarray}
\end{widetext}
Now clearly for nearest-neighbor coupling the above reduces to Eq.~(4). This result can also be generalized to higher dimensions by simply treating all dummy indices as vectors (in position or momentum space). For long-range coupling an effective coupling $\bar{J}=\sum_\Delta J(|\Delta|)|\Delta|$ can be defined to reproduce the diffusion coefficient of a nearest-neighbor coupled lattice, which, as forecasted in the main text, is dimension independent. 

\subsection{Heuristic derivation of Eq.~(6)}
Here we present a heuristic derivation of Eq.~(6) ($D_\mathrm{coh}=\Gamma\xi^2$) in the weak coupling (Redfield) regime for 1D systems, where $\xi$ is the localization length of the system at a given disorder strength. This expression was proposed on the ground of scaling arguments and verified numerically in our previous report (Ref.~6 in the main text). We start from two basic notions of quantum dynamics in disordered media in the weak coupling limit: Firstly the dynamics is adequately described by hopping rates among eigenstates, according to Eq.~(\ref{eqn:RedfieldEqn}), where the transfer rate between states is given by Eq.~(\ref{eqn:RedfieldOffDiag}). Second, a mean position $x_\mu=\langle \mu|x|\mu\rangle$ can be defined for each of the eigenstates. We argue that, again based on scaling and dimensional analysis, the diffusion coefficient is given by the following expression:
\begin{eqnarray}
D^\mathrm{1D}=\frac{1}{N}\sum_{\mu,\nu} W_{\mu\nu}x_{\mu\nu}^2,
\label{eqn:RedfieldHeuristic}
\end{eqnarray}
where $x_{\mu\nu}=x_\mu-x_\nu$ is the separation of the mean positions, and $W_{\mu\nu}$ is the transfer rate between states $\mu$ and $\nu$. To proceed we also need to specify the form of the wavefunctions. Here we assume a simple exponential decay ansatz: 
\begin{eqnarray}
\langle x|\mu\rangle = \frac{1}{\sqrt{\xi}}e^{-|x-x_\mu|/\xi}.
\end{eqnarray}
With this, the Redfield rates can be obtained as
\begin{eqnarray}
W_{\mu\nu}=\frac{\Gamma}{\xi^2}\left(|x_{\mu\nu}|+\frac{\xi}{2}\right)e^{-2|x_{\mu\nu}|/\xi}.
\end{eqnarray}
And the average in Eq.~(\ref{eqn:RedfieldHeuristic}) can be carried out straightforwardly.
\begin{eqnarray}
D^\mathrm{1D}=\frac{\Gamma}{\xi^2}\int_{-\infty}^\infty dx x^2\left(|x|+\frac{\xi}{2}\right)e^{-2|x|/\xi}=\Gamma\xi^2.
\end{eqnarray}

\subsection{Radius dependence at finite $\Gamma$}
\begin{figure}
	\centering
  \includegraphics[width=8cm]{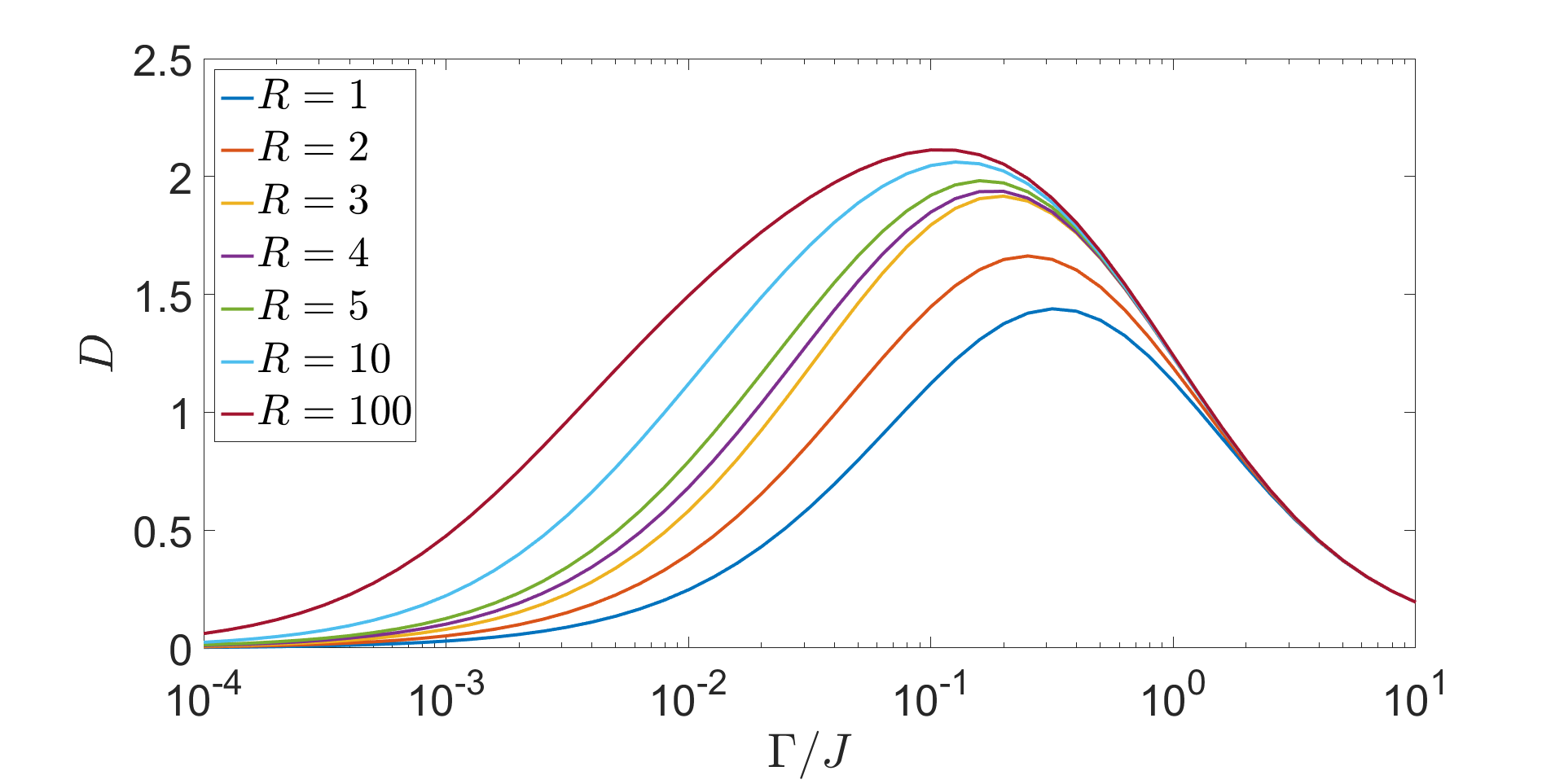}
  \caption{(Color online) $\Gamma$ dependence of $D$ at different $R$ with $\sigma/J=1$.}
  \label{fig:DofRofG}
\end{figure}

\begin{figure}
	\centering
  \includegraphics[width=8cm]{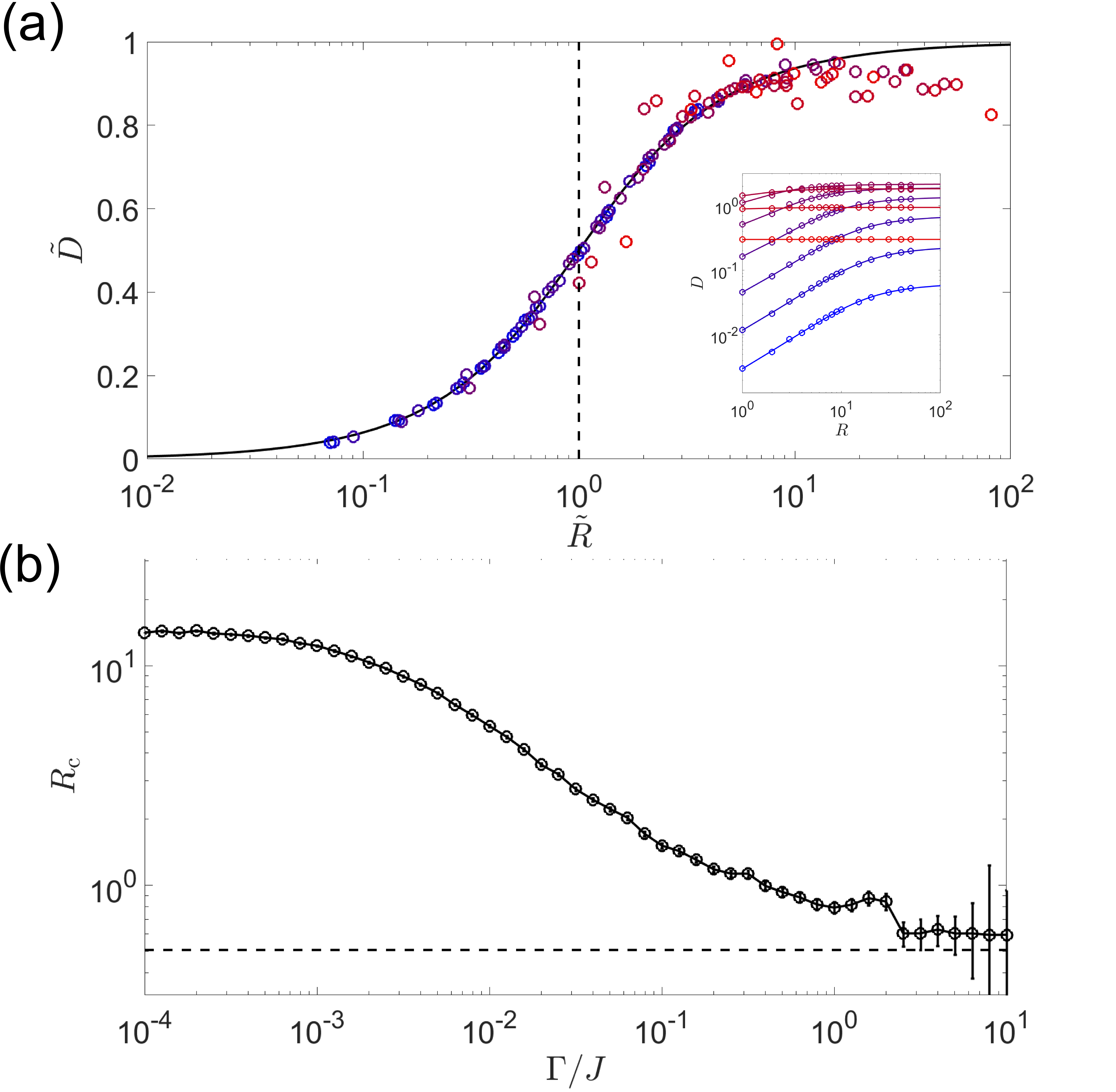}
  \caption{(Color online) (a) Relative diffusion coefficient as a function of the rescaled radius, with the convention following Fig.~2(b) in the main text. Data are taken from $\Gamma/J=10^{-4}$ (blue) to $10^1$ (red) with $\sigma/J=1$. The inset shows the original data before rescaling. (b) $\Gamma$ dependence of $R_\mathrm{c}$ fitted from Eq.~(10) with $\sigma/J=1$. The dashed line is the asymptotic value of $R_\mathrm{c}$ obtained by minimizing Eq.~(\ref{eqn:err}). The error bars show the range of standard deviation, obtained by standard bootstrap Monte Carlo sampling within the original raw data set.}
  \label{fig:RcofG}
\end{figure}

We study the radius dependence of the diffusion coefficient by focusing on the Redfield ($\Gamma/J\ll1$) regime in the main text, since the dynamics of the quantum particle in question becomes independent of the tube radius when $\Gamma/J$ is large. This is presented in Fig.~\ref{fig:DofRofG}, while the universal scaling relation with changing $\Gamma/J$ is shown in Fig.~\ref{fig:RcofG}(a). It can be seen that the general trend of interpolating $D$ between $D^\mathrm{1D}$ and $D^\mathrm{2D}$ as a function of $R$ is similar. The fitted critical radius $R_\mathrm{c}=R_\mathrm{c}(\Gamma)$ is a monotonic decreasing function of $\Gamma$ (Fig.~\ref{fig:RcofG}(b)). We point out that the fitted $R_\mathrm{c}$ is noisy at large $\Gamma/J$ since very little $R$ dependence can be recorded in that regime, represented by the red lines in the inset of Fig.~\ref{fig:RcofG}(a).


The dependence of $R_\mathrm{c}$ on $\Gamma$ can be understood as follows. In the weak system-environment coupling regime, the static disorder dominates in determining the localization length and thus the critical radius. At higher coupling strength both static disorder and dynamical noise contribute to the localization of the wavefunction. In accordance to the physical picture discussed in the main text, we expect $R_\mathrm{c}\approx\xi^\mathrm{2D}(\sigma,\Gamma)$. The localization length monotonically decreases as $\Gamma$ increases, and eventually the classical hopping mechanism takes over and the dependence on dimensionality is lost. In other words, the 2D enhancement of quantum diffusion is attenuated by the decoherence from strong thermal noise. 

Upon closer inspection of Fig.~\ref{fig:DofRofG} one finds the majority of the difference between $D^\mathrm{1D}$ and $D^\mathrm{2D}$ is manifested in the difference between $D^{R=3}$ and $D^\mathrm{1D}$ at intermediate $\Gamma$ values. This is an artifact originating from the geometric model we use: From 1D to $R=2$ to $R=3$ the number of coupled nearest-neighbors increases discontinuously from two to three to four, respectively. In contrast, in all cases with $R>3$ the coupling terms of a given site remains the same (four). Thus, in the strong dephasing regime ($\Gamma/J>1$) the $R$ dependence of the diffusion constant can be described by a two-step function: $D^\mathrm{1D}<D^{R=2}<D^{R\ge3}=D^\mathrm{2D}$. The asymptotic value of $R_c$ in this limit can be obtained by minimizing the following mean squared error.
\begin{widetext}
\begin{eqnarray}
\epsilon(R_\mathrm{c})=\left[\frac{2}{\pi}\arctan\left(\frac{1}{R_\mathrm{c}}\right)-\frac{D^{R=2}-D^\mathrm{1D}}{D^\mathrm{2D}-D^\mathrm{1D}}\right]^2+\sum_{R=3}^\infty\left[\frac{2}{\pi}\arctan\left(\frac{R-1}{R_\mathrm{c}}\right)-1\right]^2.
\label{eqn:err}
\end{eqnarray}
\end{widetext}
Assuming $(D^{R=2}-D^\mathrm{1D})/(D^\mathrm{2D}-D^\mathrm{1D})=0.49$, which is obtained by averaging our data in the range of $0.5<\Gamma/J<5$ where this phenomenon is most salient, the limiting value of the critical radius is given by $\bar{R}_\mathrm{c}=0.47$, indicated by the dashed line in Fig.~\ref{fig:RcofG}(b). 

Consequently, the critical radius (2D localization length) calculated in the Redfield regime, as elaborated in the main text, is an upper bound of its actual value. However, since the dependence on dimensionality is suppressed in the large $\Gamma$ regime, the error in estimating $D(R)$ is small. 

\subsection{Quantum bath: Temperature dependence}
\begin{figure}
	\centering
  \includegraphics[width=8cm]{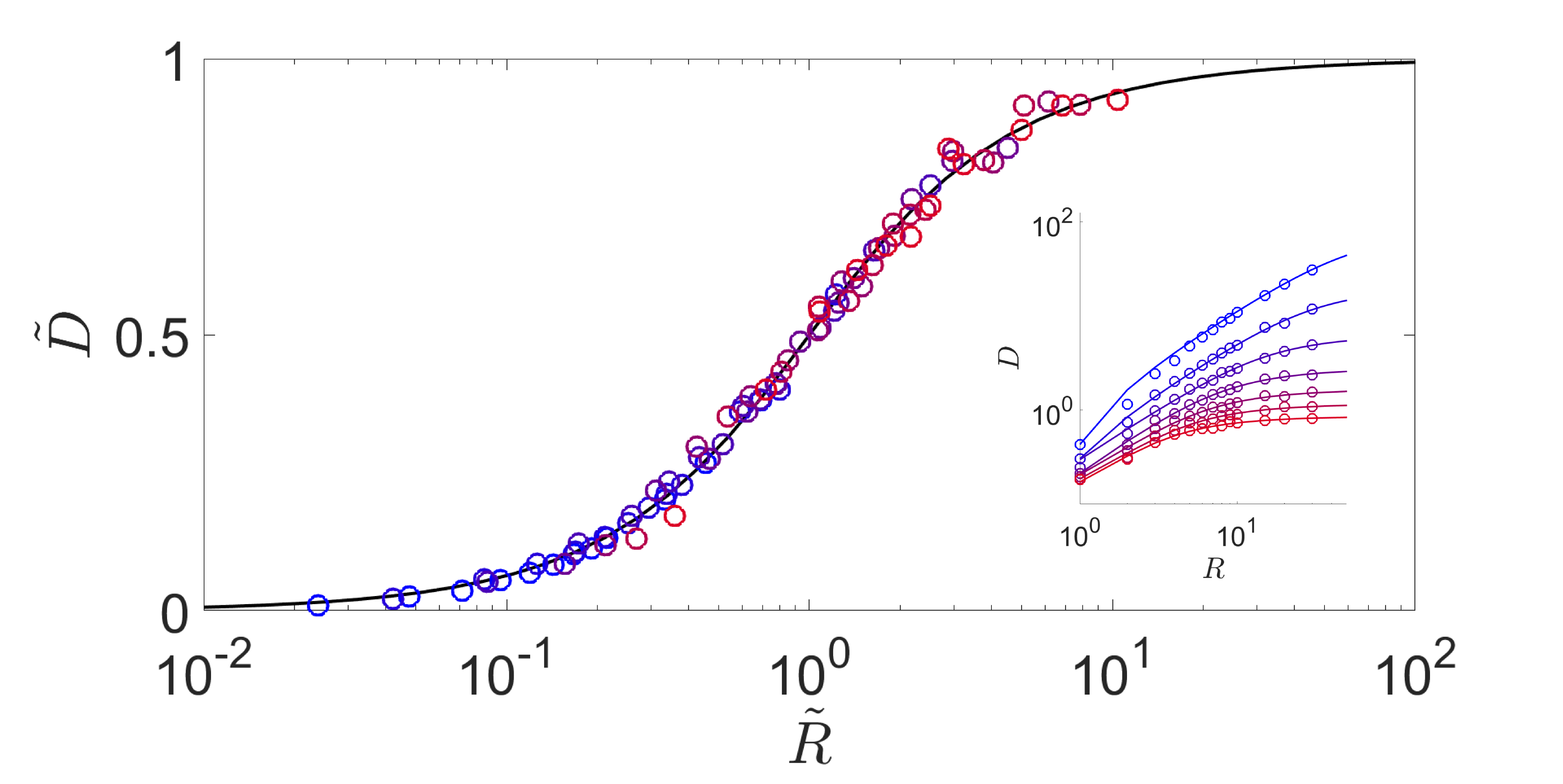}
  \caption{(Color online) Relative diffusion coefficient as a function of the rescaled radius. We use cubic super-Ohmic spectral density: $S_d(\omega)=3\lambda\Theta(\omega_\mathrm{c}-\omega)(\omega/\omega_\mathrm{c})^3$ where $\Theta(x)$ is the Heaviside function with cutoff frequency $\omega_\mathrm{c}=10J$ and the reorganization energy $\lambda=\int S_d(\omega)/\omega=0.01J$. Inset shows the original data before rescaling: From top ($\sigma/J=2$, blue) to bottom ($\sigma/J=5$, red) with $0.5$ increment and interpolating color gradient.}
  \label{fig:RcRedfield}
\end{figure}

In order to better describe the real systems listed in Table~I, where the room temperature thermal energy is comparable to or smaller than the energy gaps, we employ the secular Redfield method with a real quantum bath characterized by cubic super-Ohmic spectral density to simulate the diffusive dynamics. Under this assumption, we again obtain a master equation in the eigenbasis as in Eq.~(\ref{eqn:RedfieldEqn}), albeit with a different set of scattering rates among the states obeying detailed balance:
\begin{eqnarray}
W_{\mu\nu}&=&\sum_n|\phi_n^\mu|^2|\phi_n^\nu|^2S_d(\omega_{\mu\nu})\bar{n}(\omega_{\mu\nu},T)\label{eqn:qRedfieldOffDiag}
\end{eqnarray}
where $S_d(\omega)$ is the bath spectral density and $\bar{n}(\omega,T)=[\exp(\omega/T)-1]^{-1}$ is the Bose-Einstein distribution. We choose the temperature range to be in the thermally activated regime that the diffusivity is exponentially proportional to temperature, see Ref.~7 for details and the caption of Fig.~\ref{fig:RcRedfield} for the parameters used, which is reasonably attached to the systems in Table~I. The numerical procedure is the same as the one described in the first part of the Supplemental Material, that we first propagate the system to steady state from a localized initial state. The diffusion coefficient can then be estimated through Eq.~(\ref{eqn:RedfieldDiffusion}). Notice that Eq.~(\ref{eqn:RedfieldDiffusionHSR}) works only for the high T HSR model where forward and backward rates between states are equal in magnitude.

The universality of the radius dependence is demonstrated in Fig.~\ref{fig:RcRedfield}. Here vary the disorder strength while keeping the temperature constant, similar to that in Fig.~2(a) in the main text, where the effect of varying temperature is shown in Fig.~2(b). Obviously, different theoretical models give different estimations to the localization length that might obey different scaling laws. However, given the localization length the system length scale can be renormalized accordingly and one observes the universal radius dependence, as is explained in the main text and illustrated in Fig.~3.  

